\documentclass[10pt,aps,prd,twocolumn,showpacs,amsmath,amssymb,nofootinbib,eqsecnum,preprintnumbers,superscriptaddress]{revtex4-2}
\usepackage{amsmath,amssymb}
\usepackage{multirow}   
\usepackage{array}      

\usepackage{amsmath,amssymb}
\usepackage{enumitem}  
\usepackage[usenames, dvipsnames]{color} 
\usepackage{graphicx} 
\usepackage{comment}
\usepackage[normalem]{ulem}
\usepackage[utf8]{inputenc}
\usepackage{url}
 \usepackage{xcolor}
 \usepackage{physics}

\usepackage{hyperref}  

\usepackage{graphicx}
\usepackage{dcolumn}
\usepackage{bm}

\newcommand{\be}{\begin{equation}}
\newcommand{\ee}{\end{equation}}
\newcommand{\ba}{\begin{eqnarray}}
\newcommand{\ea}{\end{eqnarray}}

\newcommand{\beq}{\begin{equation}}
\newcommand{\eeq}{\end{equation}}
\newcommand{\beqa}{\begin{eqnarray}}
\newcommand{\eeqa}{\end{eqnarray}}

\begin{document}


\title{Charged Rotating Black Holes in Higher Dimensions}
\title{Proca-type Hair of Rotating Black Holes in Higher Dimensions}

\author{David Kubiz\v n\'ak}
\email{david.kubiznak@matfyz.cuni.cz}
\affiliation{Institute of Theoretical Physics, Faculty of Mathematics and Physics,
Charles University, Prague, V Hole{\v s}ovi{\v c}k{\' a}ch 2, 180 00 Prague 8, Czech Republic}

\author{Robert B. Mann}
\email{rbmann@uwaterloo.ca}
\affiliation{%
Department of Physics and Astronomy, University of Waterloo, 
Waterloo, ON, N2L 3G1, Canada}

\affiliation{
Perimeter Institute  for Theoretical Physics, 31 Caroline Street North, Waterloo, ON, N2L 2Y5, Canada
}%

\author{Marek Mili{\v c}ka}
\email{marek.milicka@gmail.com }
\affiliation{Institute of Theoretical Physics, Faculty of Mathematics and Physics,
Charles University, Prague, V Hole{\v s}ovi{\v c}k{\' a}ch 2, 180 00 Prague 8, Czech Republic}

\date{May 21, 2026}

\begin{abstract}
We show that spacetime symmetries 
on any background 
give rise to stealth vector fields obeying Proca-type equations supplemented by curvature terms.  This observation, which is true for solutions of any theory of gravity and with arbitrary matter content, effectively promotes spacetime symmetries to `physical fields' whose characteristic property is that their backreaction on the geometry vanishes. In particular, this  allows one to construct exact Proca hair charged and magnetized rotating black holes  in all dimensions.
In fact, such a construction is not limited to Killing vector fields and equally works for conformal Killing vectors and hidden
symmetries encoded in 
Killing--Yano tensors.
\end{abstract}

 \maketitle

\section{Introduction}
Finding {\em charged}  rotating higher-dimensional black holes is a long-standing quest that awaits completion since the groundbreaking work of Myers and Perry \cite{Myers:1986un}. Whereas in four dimensions, the Kerr--Newman solution  was obtained   almost immediately \cite{Newman:1965tw} after Kerr's discovery \cite{Kerr:1963ud}, and
in three dimensions, the charged rotating BTZ solution can be obtained by a boost of the non-rotating solution \cite{Banados:1992wn} supplemented by identifications \cite{Clement:1995zt, Martinez:1999qi}, the situation is much more complicated in higher dimensions where various no-go theorems exist, e.g. \cite{Ortaggio:2023rzp}. In fact, although several charged holes are known in higher dimensions, these are either embedded in various supergravity theories and are plagued by the presence of additional fields, e.g. \cite{Chow:2008ip, Houri:2010fr}, they are restricted to 
a slowly rotating regime \cite{Aliev:2006yk},  
or have been constructed only numerically, e.g., \cite{Kunz:2005nm}. 

A very interesting exception to this rule is the Chong--Cvetic--Lu--Pope 
solution \cite{Chong:2005hr} of the five-dimensional minimal gauged supergravity. This is a generally spinning black hole solution of  the Einstein equations coupled to the standard electromagnetic  energy-momentum tensor,  where, however, the corresponding Maxwell equations are modified by the (fine tuned) Chern--Simons self-interacting term. Interestingly, while such a construction cannot be generalized for a general coupling and to other dimensions \cite{Mir:2016dio, Blazquez-Salcedo:2016rkj}, a
Chern--Simons-like term (induced by additional non-dynamical fields) was recently used in \cite{Deshpande:2024vbn} 
to obtain charged rotating black holes in all higher dimensions, obeying the desired Einstein--Maxwell equations subject to a minimal modification of the Maxwell equation (see also \cite{Hale:2024zvu} for the 
lower-dimensional version of this procedure).

In what follows, we adopt a different strategy to tackle this problem. Namely, instead of preserving the Einstein equations and modifying the Maxwell equations, we will construct exact charged and magnetized rotating black hole solutions in a generalized Proca theory, e.g. \cite{Heisenberg:2014rta, Colladay:1998fq}, where the Maxwell equations take (at least in vacuum) the standard form, and the corresponding energy momentum tensor gets modified in such a way that it  admits {\em stealth}-like vector field configurations, 
that is, configurations characterized by a vanishing energy--momentum tensor \cite{Ayon-Beato:2004nzi}. This results in black holes that are identical to vacuum black holes in the gravitational sector but feature an `arbitrarily strong' (stealth) vector field with a non-trivial flux at infinity.

{Contrary to the numerous  (stealth) black hole solutions in generalized Proca-type theories, such as those of `vector Galileon' and Bumblebee gravity obtained in the literature so far, e.g. \cite{Chagoya:2016aar, Cisterna:2016nwq,Heisenberg:2017hwb,Heisenberg:2017xda, Xu:2022frb, Xu:2026zgd, Fernandes:2026rjs}, our solution is derived from the following general observation.}  We show that spacetime symmetries, encoded in {\em Killing vector fields},   
on any background 
give rise to stealth vector fields obeying the {\em specific}  Proca-type equations supplemented by curvature terms.  This observation, which is true for solutions of any Lagrangian-based theory of gravity  and with arbitrary matter content, effectively promotes spacetime symmetries to `physical fields' in this theory, whose characteristic property is that their backreaction on the geometry vanishes.

Our construction generalizes  the well known strategy \cite{ Papapetrou:1966zz, Wald:1974np} for obtaining {\em Maxwell test field solutions} from underlying Killing symmetries. Namely, that in any {\em vacuum} spacetime of the Einstein theory, a Killing field $\xi^a$ gives rise to a test field solution $A^a$ 
\be\label{A} 
A^a=\xi^a\,
\ee 
of the Maxwell equations: 
\be\label{MaxwellStandard} 
\nabla_a F^{ab}=0\,,\quad F=dA\,.
\ee 
This strategy, which was in particular used to construct the weakly charged Kerr black hole in an asymptotically uniform magnetic field \cite{Wald:1974np},
and its higher-dimensional generalizations  
\cite{Aliev:2004ec, Kaya:2008zz, Frolov:2010cr, Shaymatov:2024fle} (see also \cite{Aliev:2006tt, Aliev:2007qi} and \cite{Frolov:2017bdq} for its modifications), is, in fact, a special case of our construction.

Yet our construction is far more general in that: i) It allows for a construction of stealth vector fields in any background with Killing  symmetries; in that sense, it is `kinematic' (as is the definition of spacetime symmetries) and is not limited to Einstein's theory of gravity.   
ii) It promotes the weak test fields to full `backreacting'  stealth solutions with arbitrarily strong fields. 
iii) It provides a physical interpretation of spacetime symmetries, treating them as physical  `reference frames' in a given spacetime. iv) It can be straightforwardly generalized to conformal Killing fields as well as to  higher-form hidden symmetries 
\cite{us}.

We proceed as follows. We start by introducing a specific (fine tuned coupling) Proca system, whose  equations of motion coincide with the Killing vector integrability conditions and reduce to the Maxwell equations \eqref{Maxwell} in the 
vacuum Einstein case. We then show that, on account of the identification with the Killing field \eqref{A}, the corresponding energy momentum tensor vanishes, giving rise to a stealth vector solution on any background. This, in particular, allows us 
to construct rotating charged and magnetized black hole solutions with stealth vector fields in any number of spacetime dimensions. After briefly studying their properties, we conclude with possible future directions.

\section{Specific Proca theory}

{
Generalized Proca theories have a long history, e.g. \cite{Colladay:1998fq, Heisenberg:2014rta}; they encode 
Lorentz-violating extensions of the minimal standard model
\cite{Colladay:1998fq}, and can explain  the late-time acceleration of the Universe \cite{DeFelice:2016yws}.} In this paper, we focus on the  following action:
\be 
S_{\mbox{\tiny TOT}}=S_g+S_A+S_m=\int d^dx \sqrt{-g}\bigl({\cal L}_g+{\cal L}_A+{\cal L}_m\bigr)\,.
\ee  
Here, ${\cal L}_g$ describes the gravity part of the action; in Einstein gravity, this would simply be ${\cal L}_g=R/(2\kappa)$, but one can also add a cosmological constant or consider various modified gravity theories, Lovelock, for example. ${\cal L}_m$ is the other matter sector; we assume minimal coupling to gravity and no interaction with the vector $A$ sector. Finally,  we consider the vector field $A$ sector given by 
\be\label{LA} 
{\cal L}_A=-\frac{1}{4}F_{ab}F^{ab}+\beta R_{ab}A^aA^b\,,
\ee 
where, as usual, $F_{ab}=(dA)_{ab}=2\nabla_{[a}A_{b]}$, $R_{ab}$ is the Ricci tensor, and in what follows we focus on the case of a fine tuned coupling, 
\be 
\beta=1\,.
\ee 
Note that by employing the usual Stueckelberg's trick, one can introduce an `auxiliary' scalar field and promote \eqref{LA} to a gauge theory, upon which it can be coupled to a conserved current -- see Appendix~\ref{AppA} for more details. However, since this does not affect the procedure discussed below, we simply continue with \eqref{LA}.

Varying the action $S_A$ with respect to the vector field $A_a$ and with respect to the metric $g_{ab}$, we obtain the generalized vector field equations of motion
\be
\nabla_a F^{ab}+2 R^{ab}A_a=0\,,\label{Maxwell}
\ee
together with the generalized Einstein equations:
\be \label{Einstein}
G^{ab}=\kappa T^{ab}=\kappa (T^{ab}_m+T^{ab}_A)\,,
\ee 
where $G^{ab}$ is the `generalized' Einstein tensor (coming from the variation of ${\cal L}_g$), and the total energy-momentum tensor $T^{ab}$ splits into the matter, $T^{ab}_m$, and the vector field, $T^{ab}_A$, parts, $T^{ab}=T^{ab}_m+T^{ab}_A$, where 
\ba\label{TabVector} 
T^{ab}_A=T^{ab}_{\mbox{\tiny EM}}+\hat T^{ab}_A
\ea
with 
\ba 
\hat T^{ab}_A&=&g^{ab}A^cA^d R_{cd}-4A^{(a}A_c R^{b)c}-g^{ab} \nabla_c \nabla_d(A^c A^d)\nonumber\\
&&-\nabla^2 (A^aA^b)+2\nabla_c\nabla^{(a}(A^{|c|}A^{b)})\,, \label{TB} \\
T^{ab}_{\mbox{\tiny EM}}&=&F^{ac}F^b{}_c-\frac{1}{4}g^{ab} F_{cd}F^{cd}\,. 
\label{TEM}
\ea 
Obviously, in   Ricci flat spacetimes the first two terms vanish.

\section{From Killing vectors to stealth}

In what follows, we shall show that a Killing vector field $\xi$ in any background obeying the Einstein equations with arbitrary matter gives rise to a stealth solution of the above theory. Namely, we show that it obeys the generalized Maxwell equation \eqref{Maxwell} and that the corresponding  energy-momentum tensor 
\eqref{TabVector} 
vanishes:
\be\label{stealth} 
T^{ab}_\xi=0\,.
\ee 

To show this, we note that any Killing field $\xi^a$, defined by 
\be \label{KV}
\nabla_a \xi_b=\nabla_{[a}\xi_{b]}=\frac{1}{2}(d\xi)_{ab}\,,
\ee 
obeys the following identities, e.g. \cite{Wald:1984rg}:  
\ba\label{integrability} 
\nabla_a\nabla_b \xi_c&=&-\xi^d R_{dacb}\,,\quad \nabla_a (d\xi)_{bc}=-2 \xi^dR_{dacb}\,,\nonumber\\
\nabla_a \xi^a&=&0\,,\quad 
\nabla^2 \xi^a=\frac{1}{2}\nabla_b(d\xi)^{ba}=-\xi^b R^a{}_b\,.
\ea 
The last relation  immediately establishes that $A^a=\xi^a$, \eqref{A}, in any background obeys the generalized Maxwell equation \eqref{Maxwell}:
\be \label{KMax}
\nabla_b (d\xi)^{ba}+2R^a{}_b \xi^b=0\,.
\ee

To show that   the corresponding energy-momentum tensor  \eqref{TabVector}    also vanishes, we  start with 
\ba
\nabla^2(\xi^a\xi^b)&=&\nabla_c\nabla^c(\xi^a\xi^b)=
\frac{1}{2}\nabla_c\Bigl((d\xi)^{ca}\xi^b+\xi^a(d\xi)^{cb}\Bigr)\nonumber\\
&=&
-\xi^c R^a{}_c \xi^b+\frac{1}{4}(d\xi)^{ca}(d\xi)_c{}^b\nonumber\\
&&+\frac{1}{4}(d\xi)^{ca}(d\xi)_c{}^b-\xi^cR^{b}{}_c \xi^a\nonumber\\
&=&\frac{1}{2}(d\xi)^{ac}(d\xi)^b{}_c -2\xi^c R^{(a}{}_c\xi^{b)}\, .
\ea
Similarly, we find 
\ba
g^{ab}\nabla_c \nabla_d(\xi^c\xi^d)&=&g^{ab}\Bigl(R_{cd}\xi^c\xi^d-\frac{1}{4}(d\xi)^{cd}(d\xi)_{cd}\Bigr)\,,\\
\nabla_c \nabla^a(\xi^c \xi^b)&=&\xi^cR_c{}^a\xi^b+\frac{1}{4}(d\xi)^{ac}(d\xi)_c{}^b+\xi^c\nabla_c\nabla^a\xi^b\nonumber\\
&=&\xi^cR_c{}^a\xi^b-\frac{1}{4}(d\xi)^{ac}(d\xi)^b{}_c\,,
\ea  
since $\xi ^c\nabla_c\nabla^a\xi^b=-\xi^c\xi^dR_{dc}{}^{ba}=0$.
Putting everything together, we thus find that
\ba 
\hat T^{ab}_\xi&=&g^{ab}\xi^c\xi^dR_{cd}-4\xi^{(a}\xi_cR^{b)c}\nonumber\\
&&-g^{ab}\Bigl(R_{cd}\xi^c\xi^d-\frac{1}{4}(d\xi)^{cd}(d\xi)_{cd}\Bigr)\nonumber\\
&&- \frac{1}{2}(d\xi)^{ac}(d\xi)^b{}_c +2\xi^c R^{(a}{}_c\xi^{b)}\nonumber\\
&&+2\xi^cR_c{}^{(a}\xi^{b)}-\frac{1}{2} (d\xi)^{ac}(d\xi)^b{}_c\nonumber\\
&=&\frac{1}{4}g^{ab}(d\xi)^{cd}(d\xi)_{cd}-(d\xi)^{ac}(d\xi)^b{}_c\,,
\ea 
which precisely cancels the contribution from $T^{ab}_{\mbox{\tiny EM}}$ to yield the desired result \eqref{stealth}.

Thus, any solution of the Einstein equations with matter, 
\be 
G^{ab}=\kappa T^{ab}_m\,,
\ee 
can be endowed with a stealth vector $A$, given by the Killing vectors of the underlying geometry.

Let us make several remarks. 
\begin{itemize}
\item 
It is easy to verify that the `stealthiness property' of $A$ is not a simple consequence of the equations of motion \eqref{Maxwell}; in the above proof, the additional properties of Killing fields had to be used. Thus, stealth solutions generated by Killing fields are a subset of  {stealth} solutions of the above Proca theory \eqref{LA}. {Other stealth solutions, not generated by Killing fields,  also exist \cite{Fan:2017bka}.}

    \item 
The above reduces to 
the standard  
construction of test electromagnetic fields 
from underlying spacetime symmetries 
in Ricci flat spacetimes \cite{Wald:1974np, Papapetrou:1966zz}. Obviously, when $R_{ab}=0$, the additional term in the vector equation \eqref{Maxwell} vanishes, and we recover the standard Maxwell equations. 
By neglecting the backreaction, we thus recover 
a test field solution in the given background. 
The novel contribution of our construction is that now we have an exact solution of a given theory that is valid for an arbitrary field strength and in any background.

\item 
Although \eqref{LA} does not have a gauge symmetry, that is it is not invariant under 
\be \label{GI}
A\to A+d \Lambda 
\ee
 this can be remedied by employing a Stueckelberg mechanism, as shown in Appendix~\ref{AppA}.  Note that although 
the EOM \eqref{Maxwell} remain invariant
under \eqref{GI} for 
Killing vector solutions in the Ricci flat case, the stealthiness of the solution breaks and the new energy--momentum tensor no longer vanishes.

\item
Note that ${\cal L}_A$ \eqref{LA} with $\beta=1$ has the following property: when $A$ is identified with a Killing vector $\xi$, $A^a=\xi^a$:
\ba 
{\cal L}_{A=\xi}&=&-\frac{1}{2}\nabla_a \xi_b (d\xi)^{ab}+R_{ab}\xi^a\xi^b\nonumber\\
&=&
\frac{1}{2}\xi_b \nabla_a(d\xi)^{ab}+R_{ab}\xi^a\xi^b\nonumber\\
&=&-\xi_bR^b{}_a\xi^a+R_{ab}\xi^a\xi^b=0\,.
\ea
Here, we have integrated by parts, throwing away the boundary term, and used the Killing vector integrability condition \eqref{integrability}. In other words, the action vanishes (up to a boundary term) for  Killing fields.

\end{itemize}

\section{Charged and magnetized rotating black holes}
 
Let us now  turn to an important example to   illustrate 
our procedure with the construction of  
higher-dimensional rotating black holes supplemented by vector fields derived from their symmetries. We focus on the vacuum (Ricci flat) case  -- thus obtaining a  generalization of the weakly charged and magnetized Myers--Perry solutions studied in 
\cite{Wald:1974np, Aliev:2004ec, Kaya:2008zz, Frolov:2010cr, Shaymatov:2024fle}. 
The multiply-spinning Myers--Perry spacetimes in $d$ spacetime dimensions read 
\cite{Myers:1986un}:
\ba\label{HDKerr}
ds^2&=&-dt^2+\frac{2M}{U}\Bigl(dt+\sum_{i=1}^m {a_i \mu_i^2 d\phi_i}\Bigr)^2+\frac{Udr^2}{V-2M}\nonumber\\
&&+\sum_{i=1}^m{(r^2+a_i^2)}\bigl(\mu_i^2d \phi_i^2+d\mu_i^2)
+\epsilon r^2 d\nu^2\,,
\ea 
where 
\be
V=r^{\epsilon-2}\prod_{i=1}^m(r^2+a_i^2)\,,\  
U={V}\Bigl(1-\sum_{i=1}^m\frac{a_i^2\mu_i^2}{r^2+a_i^2}\Bigr)\,.
\ee
Here, $\epsilon=1, 0$ for even, odd dimensions, $m=\bigl[\frac{d-1}{2}\bigr]$ (where $[A]$ denotes the whole part of $A$), and the coordinates $\mu_i$ and $\nu$ obey a constraint 
\be 
\sum_{i=1}^m \mu_i^2+\epsilon \nu^2=1\,.
\ee

The spacetime admits the following time-like and axi-symmetric Killing fields:
\be
k=\partial_t\,,\quad \eta_i=\partial_{\phi_i}\,, 
\ee 
whose particular combination yields a Killing generator of the horizon $\xi$, 
\be 
\xi=k+\sum_{i=1}^m\Omega_i \eta_i\,,\quad \Omega_{i}=\frac{a_i}{r_+^2+a_i^2}\,.
\ee 
Here, the horizon radius $r_+$ is the largest root of $V(r_+)=2M$, and $\Omega_i$ are its 
angular velocities in respective directions.

Using these Killing vectors, we may now identify the following vector field \cite{Wald:1974np, Aliev:2004ec, Kaya:2008zz, Frolov:2010cr, Shaymatov:2024fle}
\be\label{AfullMP}
A=ck+\frac{1}{m}\sum_{i=1}^m B_i (\eta_i-a_i k)\,, 
\ee
where 
$B_i$ are the asymptotically `uniform magnetic fields'. Such a field has a non-trivial flux at infinity, given by (see Appendix~\ref{AppA}): 
\be 
{\cal F}\equiv\frac{1}{S_{d-2}}\int *F=\frac{c}{S_{d-2}}\int *(dk)=-\frac{16\pi c(d-3)}{(d-2) S_{d-2}}M\,,
\ee 
upon using the Komar definition of mass, e.g. \cite{Kastor:2009wy}, where $S_{n-1}=2 \pi^{n/2}/\Gamma(n/2)$ is the volume of a unit $(n-1)$ sphere.
Thus, choosing 
\be 
c\equiv-\frac{Q(d-2)S_{d-2}}{16\pi (d-3)M}\,,
\ee 
would correspond to a `Coulombic charge' $Q$.

In particular, setting all $B_i=0$, we have 
\be 
A=c
\Bigl[\Bigl(-1+\frac{2M}{U}\Bigr)dt+\frac{2M}{U}\sum_{i=1}^m a_i \mu_i d\phi_i\Bigr]\,,
\ee
and  
recover a stealth generalization of the weakly charged Myers--Perry black holes \cite{Wald:1974np, Aliev:2004ec, Kaya:2008zz}, with the vector field regular on the horizon. 
In 4D, this solution reduces to the one recently obtained in \cite{Xu:2026zgd}.
Interestingly, following \cite{Fan:2017bka}, we have checked that in the 4D spherical case, the above vector does not modify the thermodynamic properties of the hole, which remain identical to those of the vacuum 4D Schwarzschild solution. 
We expect that a similar property also holds in the higher-dimensional rotating case. 
Thus, although the above higher-dimensional black hole is charged in that it is supplemented by a vector field with non-trivial flux at infinity, its physical charges remain identical to those of the original vacuum solution. It remains to be seen whether this is a general feature of our stealth solutions.

\section{Discussion}

In this paper, we have shown how to promote spacetime symmetries in a given spacetime 
to stealth solutions 
 by coupling gravity to  a specific Proca-type theory \eqref{LA} with $\beta=1$.
Such a construction, which  generalizes the well known trick for obtaining test Maxwell solutions in Ricci flat spacetimes,  is purely kinematic in that it works on any background and for any  Lagrangian-based gravity theory with arbitrary matter content. In a sense, it provides a `visualization' of symmetries, which can now be treated as `physical fields' (or reference frames), obeying their own equations of motion (identified with integrability conditions of Killing fields) in the given spacetime.

In particular, this allows the construction of charged and rotating black holes in all higher-dimensions, characterized by an arbitrarily strong non-trivial flux of the stealth vector field at infinity, effectively promoting the previously constructed weakly charged black holes \cite{Wald:1974np, Aliev:2004ec, Kaya:2008zz} 
to an exact solution. While it seems that such vector hair is not directly encoded in the thermodynamic properties of the solution, it is, for example, discoverable by the motion of test particles  coupled to  this field. {It might also serve as a `seed' for generating  non-stealth solutions, similar to \cite{BenAchour:2025lkx}.}

While we have primarily focused on Killing vector fields, the above construction is much more general and can be applied to other symmetries. For example,  
consider a {\em conformal Killing vector} 
generalization of Eq.~\eqref{KV}, 
\be \label{CKVdef}
\nabla_{(a}\xi_{b)}=2\psi g_{ab}\,,\quad \psi\equiv\frac{1}{d}\nabla_a\xi^a\,,
\ee 
and generalize the action \eqref{LA} to the following:
\be\label{LACKV} 
\tilde{\cal L}_A=-\frac{1}{4}F_{ab}F^{ab}+R_{ab}A^aA^b-\frac{d-1}{d}
(\nabla_a A^a)^2\,.
\ee 
Then, by using the appropriately generalized integrability condition of \eqref{CKVdef}, it can be shown (see Appendix~\ref{AppB}) 
that any conformal Killing field obeys the corresponding Proca-type equation of motion  
\be
\nabla_a F^{ab} + 2 R^{ab}A_a+\frac{2(d-1)}{d}\nabla^b \nabla_a A^a=0\,,
\ee      
whilst yielding a stealth solution of the theory \eqref{LACKV}. In other words, conformal Killing vectors can also be promoted to stealth solutions on any background. In fact, as we will show in a follow up paper \cite{us}, this property generalizes to   Killing--Yano forms of arbitrary rank, e.g. \cite{Frolov:2017kze}, giving rise to 
higher-form stealth  solutions.

\subsection*{Acknowledgements}
We would like to thank Adolfo Cisterna and Milan Vrána for discussions. 
D.K. acknowledges support from the Charles University Research Center Grant No. UNCE24/SCI/016.

\appendix

\section{Proca gauge theory from Stueckelberg's trick}
\label{AppA}

We demonstrate that the Lagrangian \eqref{LA} can easily be promoted to a gauge theory by employing the well known Stueckelberg trick, \cite{Heisenberg:2014rta} upon which, similar to ordinary electromagnetism,  one can couple the stealth field to sources via a conserved current.  

To this end, let us introduce a real scalar field $\phi$ and modify the Lagrangian \eqref{LA} as follows:
\be
{\cal L} = -\frac{1}{4}F_{ab}F^{ab}+R_{ab}(A^a -\nabla^a \phi)(A^b-\nabla^b \phi)
+ J^a A_a\,,
\ee
where $J^a$ is a current that couples to
the vector potential $A^a$. The above Lagrangian is invariant under the gauge transformation
\be
A \to A+ d\alpha\,, \qquad \phi\to\phi + \alpha\,, 
\ee
provided $\nabla_aJ^a = 0$.
 The theory \eqref{LA} is recovered upon gauge fixing $\phi=$const. and setting $J^a=0$.

The equation of motion for $A^a$ is
\be
\nabla_a F^{ab}+2 R^{ab}(A_a -\nabla_a\phi)
= J^b\,,
\ee
and for $\phi$
\be
2\nabla_a(R^{ab}(A_b - \nabla_b\phi))=0\,,
\ee
which is exactly the divergence of the previous equation. 

Repeating the argument in the main text, it is obvious that by identifying 
\be 
A^a-\nabla^a\phi=\xi^a\,,
\ee 
for a Killing vector field $\xi^a$, we obtain a stealth solution of the above theory, with $J^a=0$; to construct the stealth solution, the vector field  $A^a$ now need only be a Killing vector up to a gauge transformation, whereas in the Proca-type theory \eqref{LA} it must be a  Killing vector.

Since  $\nabla_aJ^a = 0$, its conserved charge (by definition) is given by 
\be
Q= \int_{\Sigma} d\sigma_a J^a \,,  
\ee
where  $d\sigma_a(x)$ 
is the volume element of the spacelike hypersurface $\Sigma$. By using  $\nabla_aJ^a = 0$ and suitable boundary conditions on the current density, 
it is straightforward to show that such $Q$ is conserved in that the integral is independent of the choice of the hypersurface $\Sigma$. 
Using the above equation of motion, we then have
\begin{align}
    Q &= \int_\Sigma d\sigma_a J^a \nonumber\\
&= \int_\Sigma d\sigma_b \Bigl[\nabla_a F^{ab} + 2 R^{ab}(A_a -\nabla_a\phi)\Bigr] \nonumber\\
&= \int_{\partial \Sigma} *F 
+ 2 \int_\Sigma d\sigma_b  R^{ab}(A_a -\nabla_a\phi)\,, 
\end{align}
where $\partial \Sigma$ is the boundary of $\Sigma$.  The first term is the same as what we get from Maxwell's equations -- the charge in a given volume is given by an  integral of the field strength over the surface bounding that volume.  The second term is a volume integral, and it means that the charge of the system depends on the behavior of the gravitational field inside the volume.  Note that while the integrand of this second term is divergence-less,  it cannot be written 
as a surface integral over physical fields. For singular black holes, the second term may contribute even for   Ricci flat spacetimes. Note also that on behalf of the (last) Killing vector integrability condition \eqref{integrability}, the current $J^a$ vanishes for Killing fields (apart, potentially, at the singularity).

 To avoid the above issues with   conserved charge, in the main text we adopt a simple strategy, characterizing  our Proca-type fields with a field strength flux at infinity, namely 
\be 
{\cal F}\equiv\frac{1}{S_{d-2}}\int_{S^{d-2}_\infty} *F\,.
\ee 
As discussed in the main text, for the Myers--Perry spacetimes and for the particular choice of the Killing vector \eqref{AfullMP} this yields a non-trivial flux at infinity, given by the integration constant $Q$. Such a definition is useful as potential test particles, coupled to the vector field $A^a$, would be subject to a Lorentz-type force governed by the field strength $F^{ab}$.

\section{stealth solutions from conformal Killing vectors}
\label{AppB}

As indicated in the discussion section of the main text, the construction of stealth vector solutions also works for conformal Killing vectors. In this appendix, we prove this statement.

A conformal Killing vector $\xi^a$ obeys
\be \label{CKVdef}
\nabla_{(a}\xi_{b)}=2\psi g_{ab}\,,\quad \psi=\frac{1}{d}\nabla_a\xi^a
\ee 
and the Ricci identities then yield the following integrability condition:
\be 
\nabla_a\nabla_b \xi_c=\xi_dR^d{}_{abc}+\nabla_a(\psi g_{bc})+\nabla_b(\psi g_{ac})-\nabla_c(\psi g_{ab})\,,
\ee  
and thence also
\be 
\frac{1}{2}\nabla_a (d\xi)_{bc}=\xi^d R_{dabc}-2 g_{a[b}\nabla_{c]}\psi\,.
\ee
Contracting $ab$ indices and using the definition of $\psi$, we then have
\ba\label{CKVEOM}
\frac{1}{2}\nabla_a (d\xi)^a{}_{c}=-\xi^d R_{dc}-\frac{d-1}{d}\nabla_{c}\nabla_d\xi^d\,.
\ea
Guided by the previous result, where the equation of motion coincided with the contracted integrability condition, we may consider the following generalization of the action \eqref{LA}, namely:
\be 
\tilde{\cal L}_A=-\frac{1}{4}F_{ab}F^{ab}+R_{ab}A^aA^b-\frac{d-1}{d}
(\nabla_a A^a)^2\,,
\ee 
where we added the last term which is akin to the gauge fixing term in the Maxwell action.\footnote{
This generalization is quite natural from the following perspective. Consider Maxwell's theory, supplemented by two additional terms: 
\be\label{generalization}
\mathcal{L} = -\tfrac{1}{4}F_{ab}F^{ab} + \alpha (\nabla\cdot A)^2+\beta R_{ab}A^a A^b\,,
\ee
 where the second term is a  `gauge-fixing' term, and the last term is a `commutator term' innate to curved space. By standard manipulations, and discarding boundary terms, we then have:
\ba
\mathcal{L} &=& -\tfrac{1}{2}\nabla_a A_b F^{ab} -\alpha A^a\nabla_a\nabla_b A^b+\beta R_{ab}A^a A^b\nonumber\\
&=&  -\tfrac{1}{2}\nabla_a A_b (\nabla^a A^b - \nabla^b A^a) -\alpha A^a\nabla_a\nabla_b A^b+\beta R_{ab}A^a A^b\nonumber\\
&=& \tfrac{1}{2} A^b(\nabla^2 A_b -\nabla_a\nabla_b A^a) -\alpha A^a\nabla_a\nabla_b A^b+\beta R_{ab}A^a A^b\nonumber\\
&=&\tfrac{1}{2}A^a\Bigl[g_{ab} \nabla^2 - (1+2\alpha)\nabla_a \nabla_b + (1+2\beta) R_{ab}\Bigr]A^b\,.\nonumber\\
\ea
To obtain a `nice' propagator in curved space, one can fix $\alpha=-\tfrac{1}{2}=\beta$. 
}
The equations of motion for $A$ are 
\be
\nabla_a F^{ab} + 2 R^{ab}A_a+\frac{2(d-1)}{d}\nabla^b \nabla_a A^a=0\,.
\ee 
They are obviously satisfied for conformal Killing vectors, on behalf of the integrability condition  \eqref{CKVEOM}.

Moreover, the extra term in action contributes 
\be 
T^{ab}_A\to T^{ab}\equiv T^{ab}_A +T^{ab}_{\mbox{\tiny extra}}\,,
\ee 
to the energy-momentum tensor, where $T^{ab}_A$ is given by \eqref{TabVector}, and 
\be
T^{ab}_{\mbox{\tiny extra}}=\frac{(1-d)}{d} g^{ab} (2 A^{c} \nabla_{c}\nabla_{d}A^{d} + \nabla_{c}A^{c} \nabla_{d}A^{d})\,.
\ee
This, together with a more complicated integrability condition, makes the simplification of the complete energy-momentum tensor a bit  cumbersome. 
We start by eliminating all curvature terms employing  \eqref{CKVEOM}, and using $\nabla_aA_b = \nabla_{(a}A_{b)}+\nabla_{[a}A_{b]} \equiv{S_{ab}+V_{ab}}$, where we have denoted the symmetric part by $S$ and the antisymmetric one by $V$. This yields
\ba\label{SymDecomposition}
T_{ab}&=&\frac{1}{d}\big(4d\, V^{c(b} S^{a)}{}_{c} -2d\, S^{ab} S^{c}{}_{c} +  d\,g^{ab} S_{cd} S^{cd}\nonumber\\
&&+  g^{ab} S^{c}{}_{c}S^{d}{}_{d} - 4d\, A^{c} \nabla_{c}S^{ab} + 2 A^{c} g^{ab} \nabla_{c}S^{d}{}_{d}\nonumber\\
&&- 2d\, A^{c} g^{ab} \nabla_{d}S_{c}{}^{d}\big)\,.
\ea
Note that for $S_{ab}=0$, as is the case for Killing vectors, the expression vanishes. For the conformal Killing vectors, we can replace $S_{ab}$ using \eqref{CKVdef}, which gives a vanishing energy-momentum tensor, or in other words, a stealth solution.


\begin{thebibliography}{42}%
\makeatletter
\providecommand \@ifxundefined [1]{%
 \@ifx{#1\undefined}
}%
\providecommand \@ifnum [1]{%
 \ifnum #1\expandafter \@firstoftwo
 \else \expandafter \@secondoftwo
 \fi
}%
\providecommand \@ifx [1]{%
 \ifx #1\expandafter \@firstoftwo
 \else \expandafter \@secondoftwo
 \fi
}%
\providecommand \natexlab [1]{#1}%
\providecommand \enquote  [1]{``#1''}%
\providecommand \bibnamefont  [1]{#1}%
\providecommand \bibfnamefont [1]{#1}%
\providecommand \citenamefont [1]{#1}%
\providecommand \href@noop [0]{\@secondoftwo}%
\providecommand \href [0]{\begingroup \@sanitize@url \@href}%
\providecommand \@href[1]{\@@startlink{#1}\@@href}%
\providecommand \@@href[1]{\endgroup#1\@@endlink}%
\providecommand \@sanitize@url [0]{\catcode `\\12\catcode `\$12\catcode `\&12\catcode `\#12\catcode `\^12\catcode `\_12\catcode `\%12\relax}%
\providecommand \@@startlink[1]{}%
\providecommand \@@endlink[0]{}%
\providecommand \url  [0]{\begingroup\@sanitize@url \@url }%
\providecommand \@url [1]{\endgroup\@href {#1}{\urlprefix }}%
\providecommand \urlprefix  [0]{URL }%
\providecommand \Eprint [0]{\href }%
\providecommand \doibase [0]{https://doi.org/}%
\providecommand \selectlanguage [0]{\@gobble}%
\providecommand \bibinfo  [0]{\@secondoftwo}%
\providecommand \bibfield  [0]{\@secondoftwo}%
\providecommand \translation [1]{[#1]}%
\providecommand \BibitemOpen [0]{}%
\providecommand \bibitemStop [0]{}%
\providecommand \bibitemNoStop [0]{.\EOS\space}%
\providecommand \EOS [0]{\spacefactor3000\relax}%
\providecommand \BibitemShut  [1]{\csname bibitem#1\endcsname}%
\let\auto@bib@innerbib\@empty
\bibitem [{\citenamefont {Myers}\ and\ \citenamefont {Perry}(1986)}]{Myers:1986un}%
  \BibitemOpen
  \bibfield  {author} {\bibinfo {author} {\bibfnamefont {R.~C.}\ \bibnamefont {Myers}}\ and\ \bibinfo {author} {\bibfnamefont {M.~J.}\ \bibnamefont {Perry}},\ }\bibfield  {title} {\bibinfo {title} {{Black Holes in Higher Dimensional Space-Times}},\ }\href {https://doi.org/10.1016/0003-4916(86)90186-7} {\bibfield  {journal} {\bibinfo  {journal} {Annals Phys.}\ }\textbf {\bibinfo {volume} {172}},\ \bibinfo {pages} {304} (\bibinfo {year} {1986})}\BibitemShut {NoStop}%
\bibitem [{\citenamefont {Newman}\ and\ \citenamefont {Janis}(1965)}]{Newman:1965tw}%
  \BibitemOpen
  \bibfield  {author} {\bibinfo {author} {\bibfnamefont {E.~T.}\ \bibnamefont {Newman}}\ and\ \bibinfo {author} {\bibfnamefont {A.~I.}\ \bibnamefont {Janis}},\ }\bibfield  {title} {\bibinfo {title} {{Note on the Kerr spinning particle metric}},\ }\href {https://doi.org/10.1063/1.1704350} {\bibfield  {journal} {\bibinfo  {journal} {J. Math. Phys.}\ }\textbf {\bibinfo {volume} {6}},\ \bibinfo {pages} {915} (\bibinfo {year} {1965})}\BibitemShut {NoStop}%
\bibitem [{\citenamefont {Kerr}(1963)}]{Kerr:1963ud}%
  \BibitemOpen
  \bibfield  {author} {\bibinfo {author} {\bibfnamefont {R.~P.}\ \bibnamefont {Kerr}},\ }\bibfield  {title} {\bibinfo {title} {{Gravitational field of a spinning mass as an example of algebraically special metrics}},\ }\href {https://doi.org/10.1103/PhysRevLett.11.237} {\bibfield  {journal} {\bibinfo  {journal} {Phys. Rev. Lett.}\ }\textbf {\bibinfo {volume} {11}},\ \bibinfo {pages} {237} (\bibinfo {year} {1963})}\BibitemShut {NoStop}%
\bibitem [{\citenamefont {Banados}\ \emph {et~al.}(1992)\citenamefont {Banados}, \citenamefont {Teitelboim},\ and\ \citenamefont {Zanelli}}]{Banados:1992wn}%
  \BibitemOpen
  \bibfield  {author} {\bibinfo {author} {\bibfnamefont {M.}~\bibnamefont {Banados}}, \bibinfo {author} {\bibfnamefont {C.}~\bibnamefont {Teitelboim}},\ and\ \bibinfo {author} {\bibfnamefont {J.}~\bibnamefont {Zanelli}},\ }\bibfield  {title} {\bibinfo {title} {{The Black hole in three-dimensional space-time}},\ }\href {https://doi.org/10.1103/PhysRevLett.69.1849} {\bibfield  {journal} {\bibinfo  {journal} {Phys. Rev. Lett.}\ }\textbf {\bibinfo {volume} {69}},\ \bibinfo {pages} {1849} (\bibinfo {year} {1992})},\ \Eprint {https://arxiv.org/abs/hep-th/9204099} {arXiv:hep-th/9204099} \BibitemShut {NoStop}%
\bibitem [{\citenamefont {Clement}(1996)}]{Clement:1995zt}%
  \BibitemOpen
  \bibfield  {author} {\bibinfo {author} {\bibfnamefont {G.}~\bibnamefont {Clement}},\ }\bibfield  {title} {\bibinfo {title} {{Spinning charged BTZ black holes and selfdual particle - like solutions}},\ }\href {https://doi.org/10.1016/0370-2693(95)01464-0} {\bibfield  {journal} {\bibinfo  {journal} {Phys. Lett. B}\ }\textbf {\bibinfo {volume} {367}},\ \bibinfo {pages} {70} (\bibinfo {year} {1996})},\ \Eprint {https://arxiv.org/abs/gr-qc/9510025} {arXiv:gr-qc/9510025} \BibitemShut {NoStop}%
\bibitem [{\citenamefont {Martinez}\ \emph {et~al.}(2000)\citenamefont {Martinez}, \citenamefont {Teitelboim},\ and\ \citenamefont {Zanelli}}]{Martinez:1999qi}%
  \BibitemOpen
  \bibfield  {author} {\bibinfo {author} {\bibfnamefont {C.}~\bibnamefont {Martinez}}, \bibinfo {author} {\bibfnamefont {C.}~\bibnamefont {Teitelboim}},\ and\ \bibinfo {author} {\bibfnamefont {J.}~\bibnamefont {Zanelli}},\ }\bibfield  {title} {\bibinfo {title} {{Charged rotating black hole in three space-time dimensions}},\ }\href {https://doi.org/10.1103/PhysRevD.61.104013} {\bibfield  {journal} {\bibinfo  {journal} {Phys. Rev. D}\ }\textbf {\bibinfo {volume} {61}},\ \bibinfo {pages} {104013} (\bibinfo {year} {2000})},\ \Eprint {https://arxiv.org/abs/hep-th/9912259} {arXiv:hep-th/9912259} \BibitemShut {NoStop}%
\bibitem [{\citenamefont {Ortaggio}\ and\ \citenamefont {Srinivasan}(2024)}]{Ortaggio:2023rzp}%
  \BibitemOpen
  \bibfield  {author} {\bibinfo {author} {\bibfnamefont {M.}~\bibnamefont {Ortaggio}}\ and\ \bibinfo {author} {\bibfnamefont {A.}~\bibnamefont {Srinivasan}},\ }\bibfield  {title} {\bibinfo {title} {{Charging Kerr-Schild spacetimes in higher dimensions}},\ }\href {https://doi.org/10.1103/PhysRevD.110.044035} {\bibfield  {journal} {\bibinfo  {journal} {Phys. Rev. D}\ }\textbf {\bibinfo {volume} {110}},\ \bibinfo {pages} {044035} (\bibinfo {year} {2024})},\ \Eprint {https://arxiv.org/abs/2309.02900} {arXiv:2309.02900 [gr-qc]} \BibitemShut {NoStop}%
\bibitem [{\citenamefont {Chow}(2010)}]{Chow:2008ip}%
  \BibitemOpen
  \bibfield  {author} {\bibinfo {author} {\bibfnamefont {D.~D.~K.}\ \bibnamefont {Chow}},\ }\bibfield  {title} {\bibinfo {title} {{Charged rotating black holes in six-dimensional gauged supergravity}},\ }\href {https://doi.org/10.1088/0264-9381/27/6/065004} {\bibfield  {journal} {\bibinfo  {journal} {Class. Quant. Grav.}\ }\textbf {\bibinfo {volume} {27}},\ \bibinfo {pages} {065004} (\bibinfo {year} {2010})},\ \Eprint {https://arxiv.org/abs/0808.2728} {arXiv:0808.2728 [hep-th]} \BibitemShut {NoStop}%
\bibitem [{\citenamefont {Houri}\ \emph {et~al.}(2010)\citenamefont {Houri}, \citenamefont {Kubiznak}, \citenamefont {Warnick},\ and\ \citenamefont {Yasui}}]{Houri:2010fr}%
  \BibitemOpen
  \bibfield  {author} {\bibinfo {author} {\bibfnamefont {T.}~\bibnamefont {Houri}}, \bibinfo {author} {\bibfnamefont {D.}~\bibnamefont {Kubiznak}}, \bibinfo {author} {\bibfnamefont {C.~M.}\ \bibnamefont {Warnick}},\ and\ \bibinfo {author} {\bibfnamefont {Y.}~\bibnamefont {Yasui}},\ }\bibfield  {title} {\bibinfo {title} {{Generalized hidden symmetries and the Kerr-Sen black hole}},\ }\href {https://doi.org/10.1007/JHEP07(2010)055} {\bibfield  {journal} {\bibinfo  {journal} {JHEP}\ }\textbf {\bibinfo {volume} {07}},\ \bibinfo {pages} {055}},\ \Eprint {https://arxiv.org/abs/1004.1032} {arXiv:1004.1032 [hep-th]} \BibitemShut {NoStop}%
\bibitem [{\citenamefont {Aliev}(2006)}]{Aliev:2006yk}%
  \BibitemOpen
  \bibfield  {author} {\bibinfo {author} {\bibfnamefont {A.~N.}\ \bibnamefont {Aliev}},\ }\bibfield  {title} {\bibinfo {title} {{Rotating black holes in higher dimensional Einstein-Maxwell gravity}},\ }\href {https://doi.org/10.1103/PhysRevD.74.024011} {\bibfield  {journal} {\bibinfo  {journal} {Phys. Rev. D}\ }\textbf {\bibinfo {volume} {74}},\ \bibinfo {pages} {024011} (\bibinfo {year} {2006})},\ \Eprint {https://arxiv.org/abs/hep-th/0604207} {arXiv:hep-th/0604207} \BibitemShut {NoStop}%
\bibitem [{\citenamefont {Kunz}\ \emph {et~al.}(2005)\citenamefont {Kunz}, \citenamefont {Navarro-Lerida},\ and\ \citenamefont {Petersen}}]{Kunz:2005nm}%
  \BibitemOpen
  \bibfield  {author} {\bibinfo {author} {\bibfnamefont {J.}~\bibnamefont {Kunz}}, \bibinfo {author} {\bibfnamefont {F.}~\bibnamefont {Navarro-Lerida}},\ and\ \bibinfo {author} {\bibfnamefont {A.~K.}\ \bibnamefont {Petersen}},\ }\bibfield  {title} {\bibinfo {title} {{Five-dimensional charged rotating black holes}},\ }\href {https://doi.org/10.1016/j.physletb.2005.03.056} {\bibfield  {journal} {\bibinfo  {journal} {Phys. Lett. B}\ }\textbf {\bibinfo {volume} {614}},\ \bibinfo {pages} {104} (\bibinfo {year} {2005})},\ \Eprint {https://arxiv.org/abs/gr-qc/0503010} {arXiv:gr-qc/0503010} \BibitemShut {NoStop}%
\bibitem [{\citenamefont {Chong}\ \emph {et~al.}(2005)\citenamefont {Chong}, \citenamefont {Cvetic}, \citenamefont {Lu},\ and\ \citenamefont {Pope}}]{Chong:2005hr}%
  \BibitemOpen
  \bibfield  {author} {\bibinfo {author} {\bibfnamefont {Z.~W.}\ \bibnamefont {Chong}}, \bibinfo {author} {\bibfnamefont {M.}~\bibnamefont {Cvetic}}, \bibinfo {author} {\bibfnamefont {H.}~\bibnamefont {Lu}},\ and\ \bibinfo {author} {\bibfnamefont {C.~N.}\ \bibnamefont {Pope}},\ }\bibfield  {title} {\bibinfo {title} {{General non-extremal rotating black holes in minimal five-dimensional gauged supergravity}},\ }\href {https://doi.org/10.1103/PhysRevLett.95.161301} {\bibfield  {journal} {\bibinfo  {journal} {Phys. Rev. Lett.}\ }\textbf {\bibinfo {volume} {95}},\ \bibinfo {pages} {161301} (\bibinfo {year} {2005})},\ \Eprint {https://arxiv.org/abs/hep-th/0506029} {arXiv:hep-th/0506029} \BibitemShut {NoStop}%
\bibitem [{\citenamefont {Mir}\ and\ \citenamefont {Mann}(2017)}]{Mir:2016dio}%
  \BibitemOpen
  \bibfield  {author} {\bibinfo {author} {\bibfnamefont {M.}~\bibnamefont {Mir}}\ and\ \bibinfo {author} {\bibfnamefont {R.~B.}\ \bibnamefont {Mann}},\ }\bibfield  {title} {\bibinfo {title} {{Charged Rotating AdS Black Holes with Chern-Simons coupling}},\ }\href {https://doi.org/10.1103/PhysRevD.95.024005} {\bibfield  {journal} {\bibinfo  {journal} {Phys. Rev. D}\ }\textbf {\bibinfo {volume} {95}},\ \bibinfo {pages} {024005} (\bibinfo {year} {2017})},\ \Eprint {https://arxiv.org/abs/1610.05281} {arXiv:1610.05281 [gr-qc]} \BibitemShut {NoStop}%
\bibitem [{\citenamefont {Bl{\'a}zquez-Salcedo}\ \emph {et~al.}(2017)\citenamefont {Bl{\'a}zquez-Salcedo}, \citenamefont {Kunz}, \citenamefont {Navarro-L{\'e}rida},\ and\ \citenamefont {Radu}}]{Blazquez-Salcedo:2016rkj}%
  \BibitemOpen
  \bibfield  {author} {\bibinfo {author} {\bibfnamefont {J.~L.}\ \bibnamefont {Bl{\'a}zquez-Salcedo}}, \bibinfo {author} {\bibfnamefont {J.}~\bibnamefont {Kunz}}, \bibinfo {author} {\bibfnamefont {F.}~\bibnamefont {Navarro-L{\'e}rida}},\ and\ \bibinfo {author} {\bibfnamefont {E.}~\bibnamefont {Radu}},\ }\bibfield  {title} {\bibinfo {title} {{Charged rotating black holes in Einstein-Maxwell-Chern-Simons theory with a negative cosmological constant}},\ }\href {https://doi.org/10.1103/PhysRevD.95.064018} {\bibfield  {journal} {\bibinfo  {journal} {Phys. Rev. D}\ }\textbf {\bibinfo {volume} {95}},\ \bibinfo {pages} {064018} (\bibinfo {year} {2017})},\ \Eprint {https://arxiv.org/abs/1610.05282} {arXiv:1610.05282 [gr-qc]} \BibitemShut {NoStop}%
\bibitem [{\citenamefont {Deshpande}\ and\ \citenamefont {Lunin}(2025)}]{Deshpande:2024vbn}%
  \BibitemOpen
  \bibfield  {author} {\bibinfo {author} {\bibfnamefont {R.}~\bibnamefont {Deshpande}}\ and\ \bibinfo {author} {\bibfnamefont {O.}~\bibnamefont {Lunin}},\ }\bibfield  {title} {\bibinfo {title} {{Rotating Einstein-Maxwell black holes in odd dimensions}},\ }\href {https://doi.org/10.1007/JHEP06(2025)066} {\bibfield  {journal} {\bibinfo  {journal} {JHEP}\ }\textbf {\bibinfo {volume} {06}},\ \bibinfo {pages} {066}},\ \Eprint {https://arxiv.org/abs/2411.01795} {arXiv:2411.01795 [hep-th]} \BibitemShut {NoStop}%
\bibitem [{\citenamefont {Hale}\ \emph {et~al.}(2025)\citenamefont {Hale}, \citenamefont {Hull}, \citenamefont {Kubiz{\v{n}}{\'a}k}, \citenamefont {Mann},\ and\ \citenamefont {Men{\v{s}}{\'\i}kov{\'a}}}]{Hale:2024zvu}%
  \BibitemOpen
  \bibfield  {author} {\bibinfo {author} {\bibfnamefont {T.}~\bibnamefont {Hale}}, \bibinfo {author} {\bibfnamefont {B.~R.}\ \bibnamefont {Hull}}, \bibinfo {author} {\bibfnamefont {D.}~\bibnamefont {Kubiz{\v{n}}{\'a}k}}, \bibinfo {author} {\bibfnamefont {R.~B.}\ \bibnamefont {Mann}},\ and\ \bibinfo {author} {\bibfnamefont {J.}~\bibnamefont {Men{\v{s}}{\'\i}kov{\'a}}},\ }\bibfield  {title} {\bibinfo {title} {{New interpretation of the original charged BTZ black hole spacetime}},\ }\href {https://doi.org/10.1088/1361-6382/adc9f2} {\bibfield  {journal} {\bibinfo  {journal} {Class. Quant. Grav.}\ }\textbf {\bibinfo {volume} {42}},\ \bibinfo {pages} {09LT01} (\bibinfo {year} {2025})},\ \Eprint {https://arxiv.org/abs/2412.04329} {arXiv:2412.04329 [gr-qc]} \BibitemShut {NoStop}%
\bibitem [{\citenamefont {Heisenberg}(2014)}]{Heisenberg:2014rta}%
  \BibitemOpen
  \bibfield  {author} {\bibinfo {author} {\bibfnamefont {L.}~\bibnamefont {Heisenberg}},\ }\bibfield  {title} {\bibinfo {title} {{Generalization of the Proca Action}},\ }\href {https://doi.org/10.1088/1475-7516/2014/05/015} {\bibfield  {journal} {\bibinfo  {journal} {JCAP}\ }\textbf {\bibinfo {volume} {05}},\ \bibinfo {pages} {015}},\ \Eprint {https://arxiv.org/abs/1402.7026} {arXiv:1402.7026 [hep-th]} \BibitemShut {NoStop}%
\bibitem [{\citenamefont {Colladay}\ and\ \citenamefont {Kostelecky}(1998)}]{Colladay:1998fq}%
  \BibitemOpen
  \bibfield  {author} {\bibinfo {author} {\bibfnamefont {D.}~\bibnamefont {Colladay}}\ and\ \bibinfo {author} {\bibfnamefont {V.~A.}\ \bibnamefont {Kostelecky}},\ }\bibfield  {title} {\bibinfo {title} {{Lorentz violating extension of the standard model}},\ }\href {https://doi.org/10.1103/PhysRevD.58.116002} {\bibfield  {journal} {\bibinfo  {journal} {Phys. Rev. D}\ }\textbf {\bibinfo {volume} {58}},\ \bibinfo {pages} {116002} (\bibinfo {year} {1998})},\ \Eprint {https://arxiv.org/abs/hep-ph/9809521} {arXiv:hep-ph/9809521} \BibitemShut {NoStop}%
\bibitem [{\citenamefont {Ayon-Beato}\ \emph {et~al.}(2006)\citenamefont {Ayon-Beato}, \citenamefont {Martinez},\ and\ \citenamefont {Zanelli}}]{Ayon-Beato:2004nzi}%
  \BibitemOpen
  \bibfield  {author} {\bibinfo {author} {\bibfnamefont {E.}~\bibnamefont {Ayon-Beato}}, \bibinfo {author} {\bibfnamefont {C.}~\bibnamefont {Martinez}},\ and\ \bibinfo {author} {\bibfnamefont {J.}~\bibnamefont {Zanelli}},\ }\bibfield  {title} {\bibinfo {title} {{Stealth scalar field overflying a (2+1) black hole}},\ }\href {https://doi.org/10.1007/s10714-005-0213-x} {\bibfield  {journal} {\bibinfo  {journal} {Gen. Rel. Grav.}\ }\textbf {\bibinfo {volume} {38}},\ \bibinfo {pages} {145} (\bibinfo {year} {2006})},\ \Eprint {https://arxiv.org/abs/hep-th/0403228} {arXiv:hep-th/0403228} \BibitemShut {NoStop}%
\bibitem [{\citenamefont {Chagoya}\ \emph {et~al.}(2016)\citenamefont {Chagoya}, \citenamefont {Niz},\ and\ \citenamefont {Tasinato}}]{Chagoya:2016aar}%
  \BibitemOpen
  \bibfield  {author} {\bibinfo {author} {\bibfnamefont {J.}~\bibnamefont {Chagoya}}, \bibinfo {author} {\bibfnamefont {G.}~\bibnamefont {Niz}},\ and\ \bibinfo {author} {\bibfnamefont {G.}~\bibnamefont {Tasinato}},\ }\bibfield  {title} {\bibinfo {title} {{Black Holes and Abelian Symmetry Breaking}},\ }\href {https://doi.org/10.1088/0264-9381/33/17/175007} {\bibfield  {journal} {\bibinfo  {journal} {Class. Quant. Grav.}\ }\textbf {\bibinfo {volume} {33}},\ \bibinfo {pages} {175007} (\bibinfo {year} {2016})},\ \Eprint {https://arxiv.org/abs/1602.08697} {arXiv:1602.08697 [hep-th]} \BibitemShut {NoStop}%
\bibitem [{\citenamefont {Cisterna}\ \emph {et~al.}(2016)\citenamefont {Cisterna}, \citenamefont {Hassaine}, \citenamefont {Oliva},\ and\ \citenamefont {Rinaldi}}]{Cisterna:2016nwq}%
  \BibitemOpen
  \bibfield  {author} {\bibinfo {author} {\bibfnamefont {A.}~\bibnamefont {Cisterna}}, \bibinfo {author} {\bibfnamefont {M.}~\bibnamefont {Hassaine}}, \bibinfo {author} {\bibfnamefont {J.}~\bibnamefont {Oliva}},\ and\ \bibinfo {author} {\bibfnamefont {M.}~\bibnamefont {Rinaldi}},\ }\bibfield  {title} {\bibinfo {title} {{Static and rotating solutions for Vector-Galileon theories}},\ }\href {https://doi.org/10.1103/PhysRevD.94.104039} {\bibfield  {journal} {\bibinfo  {journal} {Phys. Rev. D}\ }\textbf {\bibinfo {volume} {94}},\ \bibinfo {pages} {104039} (\bibinfo {year} {2016})},\ \Eprint {https://arxiv.org/abs/1609.03430} {arXiv:1609.03430 [gr-qc]} \BibitemShut {NoStop}%
\bibitem [{\citenamefont {Heisenberg}\ \emph {et~al.}(2017{\natexlab{a}})\citenamefont {Heisenberg}, \citenamefont {Kase}, \citenamefont {Minamitsuji},\ and\ \citenamefont {Tsujikawa}}]{Heisenberg:2017hwb}%
  \BibitemOpen
  \bibfield  {author} {\bibinfo {author} {\bibfnamefont {L.}~\bibnamefont {Heisenberg}}, \bibinfo {author} {\bibfnamefont {R.}~\bibnamefont {Kase}}, \bibinfo {author} {\bibfnamefont {M.}~\bibnamefont {Minamitsuji}},\ and\ \bibinfo {author} {\bibfnamefont {S.}~\bibnamefont {Tsujikawa}},\ }\bibfield  {title} {\bibinfo {title} {{Black holes in vector-tensor theories}},\ }\href {https://doi.org/10.1088/1475-7516/2017/08/024} {\bibfield  {journal} {\bibinfo  {journal} {JCAP}\ }\textbf {\bibinfo {volume} {08}},\ \bibinfo {pages} {024}},\ \Eprint {https://arxiv.org/abs/1706.05115} {arXiv:1706.05115 [gr-qc]} \BibitemShut {NoStop}%
\bibitem [{\citenamefont {Heisenberg}\ \emph {et~al.}(2017{\natexlab{b}})\citenamefont {Heisenberg}, \citenamefont {Kase}, \citenamefont {Minamitsuji},\ and\ \citenamefont {Tsujikawa}}]{Heisenberg:2017xda}%
  \BibitemOpen
  \bibfield  {author} {\bibinfo {author} {\bibfnamefont {L.}~\bibnamefont {Heisenberg}}, \bibinfo {author} {\bibfnamefont {R.}~\bibnamefont {Kase}}, \bibinfo {author} {\bibfnamefont {M.}~\bibnamefont {Minamitsuji}},\ and\ \bibinfo {author} {\bibfnamefont {S.}~\bibnamefont {Tsujikawa}},\ }\bibfield  {title} {\bibinfo {title} {{Hairy black-hole solutions in generalized Proca theories}},\ }\href {https://doi.org/10.1103/PhysRevD.96.084049} {\bibfield  {journal} {\bibinfo  {journal} {Phys. Rev. D}\ }\textbf {\bibinfo {volume} {96}},\ \bibinfo {pages} {084049} (\bibinfo {year} {2017}{\natexlab{b}})},\ \Eprint {https://arxiv.org/abs/1705.09662} {arXiv:1705.09662 [gr-qc]} \BibitemShut {NoStop}%
\bibitem [{\citenamefont {Xu}\ \emph {et~al.}(2023)\citenamefont {Xu}, \citenamefont {Liang},\ and\ \citenamefont {Shao}}]{Xu:2022frb}%
  \BibitemOpen
  \bibfield  {author} {\bibinfo {author} {\bibfnamefont {R.}~\bibnamefont {Xu}}, \bibinfo {author} {\bibfnamefont {D.}~\bibnamefont {Liang}},\ and\ \bibinfo {author} {\bibfnamefont {L.}~\bibnamefont {Shao}},\ }\bibfield  {title} {\bibinfo {title} {{Static spherical vacuum solutions in the bumblebee gravity model}},\ }\href {https://doi.org/10.1103/PhysRevD.107.024011} {\bibfield  {journal} {\bibinfo  {journal} {Phys. Rev. D}\ }\textbf {\bibinfo {volume} {107}},\ \bibinfo {pages} {024011} (\bibinfo {year} {2023})},\ \Eprint {https://arxiv.org/abs/2209.02209} {arXiv:2209.02209 [gr-qc]} \BibitemShut {NoStop}%
\bibitem [{\citenamefont {Xu}\ \emph {et~al.}(2026)\citenamefont {Xu}, \citenamefont {Mai},\ and\ \citenamefont {Liang}}]{Xu:2026zgd}%
  \BibitemOpen
  \bibfield  {author} {\bibinfo {author} {\bibfnamefont {R.}~\bibnamefont {Xu}}, \bibinfo {author} {\bibfnamefont {Z.-F.}\ \bibnamefont {Mai}},\ and\ \bibinfo {author} {\bibfnamefont {D.}~\bibnamefont {Liang}},\ }\bibfield  {title} {\bibinfo {title} {{The stealth Kerr solution in the bumblebee gravity}},\ }\href {https://doi.org/10.1016/j.physletb.2026.140364} {\bibfield  {journal} {\bibinfo  {journal} {Phys. Lett. B}\ }\textbf {\bibinfo {volume} {875}},\ \bibinfo {pages} {140364} (\bibinfo {year} {2026})},\ \Eprint {https://arxiv.org/abs/2601.18809} {arXiv:2601.18809 [gr-qc]} \BibitemShut {NoStop}%
\bibitem [{\citenamefont {Fernandes}(2026)}]{Fernandes:2026rjs}%
  \BibitemOpen
  \bibfield  {author} {\bibinfo {author} {\bibfnamefont {P.~G.~S.}\ \bibnamefont {Fernandes}},\ }\bibfield  {title} {\bibinfo {title} {{Exact analytic rotating black-hole solutions with primary hair}},\ }\href@noop {} {\  (\bibinfo {year} {2026})},\ \Eprint {https://arxiv.org/abs/2601.21163} {arXiv:2601.21163 [gr-qc]} \BibitemShut {NoStop}%
\bibitem [{\citenamefont {Papapetrou}(1966)}]{Papapetrou:1966zz}%
  \BibitemOpen
  \bibfield  {author} {\bibinfo {author} {\bibfnamefont {A.}~\bibnamefont {Papapetrou}},\ }\bibfield  {title} {\bibinfo {title} {{Champs gravitationnels stationnaires {\`a} sym{\'e}trie axiale}},\ }\href@noop {} {\bibfield  {journal} {\bibinfo  {journal} {Ann. Inst. H. Poincare Phys. Theor. A}\ }\textbf {\bibinfo {volume} {4}},\ \bibinfo {pages} {83} (\bibinfo {year} {1966})}\BibitemShut {NoStop}%
\bibitem [{\citenamefont {Wald}(1974)}]{Wald:1974np}%
  \BibitemOpen
  \bibfield  {author} {\bibinfo {author} {\bibfnamefont {R.~M.}\ \bibnamefont {Wald}},\ }\bibfield  {title} {\bibinfo {title} {{Black hole in a uniform magnetic field}},\ }\href {https://doi.org/10.1103/PhysRevD.10.1680} {\bibfield  {journal} {\bibinfo  {journal} {Phys. Rev. D}\ }\textbf {\bibinfo {volume} {10}},\ \bibinfo {pages} {1680} (\bibinfo {year} {1974})}\BibitemShut {NoStop}%
\bibitem [{\citenamefont {Aliev}\ and\ \citenamefont {Frolov}(2004)}]{Aliev:2004ec}%
  \BibitemOpen
  \bibfield  {author} {\bibinfo {author} {\bibfnamefont {A.~N.}\ \bibnamefont {Aliev}}\ and\ \bibinfo {author} {\bibfnamefont {V.~P.}\ \bibnamefont {Frolov}},\ }\bibfield  {title} {\bibinfo {title} {{Five-dimensional rotating black hole in a uniform magnetic field: The Gyromagnetic ratio}},\ }\href {https://doi.org/10.1103/PhysRevD.69.084022} {\bibfield  {journal} {\bibinfo  {journal} {Phys. Rev. D}\ }\textbf {\bibinfo {volume} {69}},\ \bibinfo {pages} {084022} (\bibinfo {year} {2004})},\ \Eprint {https://arxiv.org/abs/hep-th/0401095} {arXiv:hep-th/0401095} \BibitemShut {NoStop}%
\bibitem [{\citenamefont {Kaya}(2008)}]{Kaya:2008zz}%
  \BibitemOpen
  \bibfield  {author} {\bibinfo {author} {\bibfnamefont {R.}~\bibnamefont {Kaya}},\ }\bibfield  {title} {\bibinfo {title} {{Gyromagnetic ratio of higher dimensional black holes}},\ }\href {https://doi.org/10.1088/0264-9381/25/4/045004} {\bibfield  {journal} {\bibinfo  {journal} {Class. Quant. Grav.}\ }\textbf {\bibinfo {volume} {25}},\ \bibinfo {pages} {045004} (\bibinfo {year} {2008})}\BibitemShut {NoStop}%
\bibitem [{\citenamefont {Frolov}\ and\ \citenamefont {Krtous}(2011)}]{Frolov:2010cr}%
  \BibitemOpen
  \bibfield  {author} {\bibinfo {author} {\bibfnamefont {V.~P.}\ \bibnamefont {Frolov}}\ and\ \bibinfo {author} {\bibfnamefont {P.}~\bibnamefont {Krtous}},\ }\bibfield  {title} {\bibinfo {title} {{Charged particle in higher dimensional weakly charged rotating black hole spacetime}},\ }\href {https://doi.org/10.1103/PhysRevD.83.024016} {\bibfield  {journal} {\bibinfo  {journal} {Phys. Rev. D}\ }\textbf {\bibinfo {volume} {83}},\ \bibinfo {pages} {024016} (\bibinfo {year} {2011})},\ \Eprint {https://arxiv.org/abs/1010.2266} {arXiv:1010.2266 [hep-th]} \BibitemShut {NoStop}%
\bibitem [{\citenamefont {Shaymatov}(2024)}]{Shaymatov:2024fle}%
  \BibitemOpen
  \bibfield  {author} {\bibinfo {author} {\bibfnamefont {S.}~\bibnamefont {Shaymatov}},\ }\bibfield  {title} {\bibinfo {title} {{Efficiency of magnetic Penrose process in higher dimensional Myers-Perry black hole spacetimes}},\ }\href {https://doi.org/10.1103/PhysRevD.110.044042} {\bibfield  {journal} {\bibinfo  {journal} {Phys. Rev. D}\ }\textbf {\bibinfo {volume} {110}},\ \bibinfo {pages} {044042} (\bibinfo {year} {2024})},\ \Eprint {https://arxiv.org/abs/2402.02471} {arXiv:2402.02471 [gr-qc]} \BibitemShut {NoStop}%
\bibitem [{\citenamefont {Aliev}(2007{\natexlab{a}})}]{Aliev:2006tt}%
  \BibitemOpen
  \bibfield  {author} {\bibinfo {author} {\bibfnamefont {A.~N.}\ \bibnamefont {Aliev}},\ }\bibfield  {title} {\bibinfo {title} {{Gyromagnetic Ratio of Charged Kerr-Anti-de Sitter Black Holes}},\ }\href {https://doi.org/10.1088/0264-9381/24/18/008} {\bibfield  {journal} {\bibinfo  {journal} {Class. Quant. Grav.}\ }\textbf {\bibinfo {volume} {24}},\ \bibinfo {pages} {4669} (\bibinfo {year} {2007}{\natexlab{a}})},\ \Eprint {https://arxiv.org/abs/hep-th/0611205} {arXiv:hep-th/0611205} \BibitemShut {NoStop}%
\bibitem [{\citenamefont {Aliev}(2007{\natexlab{b}})}]{Aliev:2007qi}%
  \BibitemOpen
  \bibfield  {author} {\bibinfo {author} {\bibfnamefont {A.~N.}\ \bibnamefont {Aliev}},\ }\bibfield  {title} {\bibinfo {title} {{Electromagnetic Properties of Kerr-Anti-de Sitter Black Holes}},\ }\href {https://doi.org/10.1103/PhysRevD.75.084041} {\bibfield  {journal} {\bibinfo  {journal} {Phys. Rev. D}\ }\textbf {\bibinfo {volume} {75}},\ \bibinfo {pages} {084041} (\bibinfo {year} {2007}{\natexlab{b}})},\ \Eprint {https://arxiv.org/abs/hep-th/0702129} {arXiv:hep-th/0702129} \BibitemShut {NoStop}%
\bibitem [{\citenamefont {Frolov}\ \emph {et~al.}(2017{\natexlab{a}})\citenamefont {Frolov}, \citenamefont {Krtous},\ and\ \citenamefont {Kubiznak}}]{Frolov:2017bdq}%
  \BibitemOpen
  \bibfield  {author} {\bibinfo {author} {\bibfnamefont {V.~P.}\ \bibnamefont {Frolov}}, \bibinfo {author} {\bibfnamefont {P.}~\bibnamefont {Krtous}},\ and\ \bibinfo {author} {\bibfnamefont {D.}~\bibnamefont {Kubiznak}},\ }\bibfield  {title} {\bibinfo {title} {{Weakly charged generalized Kerr{\textendash}NUT{\textendash}(A)dS spacetimes}},\ }\href {https://doi.org/10.1016/j.physletb.2017.05.041} {\bibfield  {journal} {\bibinfo  {journal} {Phys. Lett. B}\ }\textbf {\bibinfo {volume} {771}},\ \bibinfo {pages} {254} (\bibinfo {year} {2017}{\natexlab{a}})},\ \Eprint {https://arxiv.org/abs/1705.00943} {arXiv:1705.00943 [gr-qc]} \BibitemShut {NoStop}%
\bibitem [{\citenamefont {Kubiznak}\ \emph {et~al.}()\citenamefont {Kubiznak}, \citenamefont {Mann},\ and\ \citenamefont {Mili{\v c}ka}}]{us}%
  \BibitemOpen
  \bibfield  {author} {\bibinfo {author} {\bibfnamefont {D.}~\bibnamefont {Kubiznak}}, \bibinfo {author} {\bibfnamefont {R.}~\bibnamefont {Mann}},\ and\ \bibinfo {author} {\bibfnamefont {M.}~\bibnamefont {Mili{\v c}ka}},\ }\bibfield  {title} {\bibinfo {title} {{From (hidden) symmetries to stealth solutions}},\ }\href@noop {} {\ }\BibitemShut {NoStop}%
\bibitem [{\citenamefont {De~Felice}\ \emph {et~al.}(2016)\citenamefont {De~Felice}, \citenamefont {Heisenberg}, \citenamefont {Kase}, \citenamefont {Mukohyama}, \citenamefont {Tsujikawa},\ and\ \citenamefont {Zhang}}]{DeFelice:2016yws}%
  \BibitemOpen
  \bibfield  {author} {\bibinfo {author} {\bibfnamefont {A.}~\bibnamefont {De~Felice}}, \bibinfo {author} {\bibfnamefont {L.}~\bibnamefont {Heisenberg}}, \bibinfo {author} {\bibfnamefont {R.}~\bibnamefont {Kase}}, \bibinfo {author} {\bibfnamefont {S.}~\bibnamefont {Mukohyama}}, \bibinfo {author} {\bibfnamefont {S.}~\bibnamefont {Tsujikawa}},\ and\ \bibinfo {author} {\bibfnamefont {Y.-l.}\ \bibnamefont {Zhang}},\ }\bibfield  {title} {\bibinfo {title} {{Cosmology in generalized Proca theories}},\ }\href {https://doi.org/10.1088/1475-7516/2016/06/048} {\bibfield  {journal} {\bibinfo  {journal} {JCAP}\ }\textbf {\bibinfo {volume} {06}},\ \bibinfo {pages} {048}},\ \Eprint {https://arxiv.org/abs/1603.05806} {arXiv:1603.05806 [gr-qc]} \BibitemShut {NoStop}%
\bibitem [{\citenamefont {Wald}(1984)}]{Wald:1984rg}%
  \BibitemOpen
  \bibfield  {author} {\bibinfo {author} {\bibfnamefont {R.~M.}\ \bibnamefont {Wald}},\ }\href {https://doi.org/10.7208/chicago/9780226870373.001.0001} {\emph {\bibinfo {title} {{General Relativity}}}}\ (\bibinfo  {publisher} {Chicago Univ. Pr.},\ \bibinfo {address} {Chicago, USA},\ \bibinfo {year} {1984})\BibitemShut {NoStop}%
\bibitem [{\citenamefont {Fan}(2018)}]{Fan:2017bka}%
  \BibitemOpen
  \bibfield  {author} {\bibinfo {author} {\bibfnamefont {Z.-Y.}\ \bibnamefont {Fan}},\ }\bibfield  {title} {\bibinfo {title} {{Black holes in vector-tensor theories and their thermodynamics}},\ }\href {https://doi.org/10.1140/epjc/s10052-018-5540-7} {\bibfield  {journal} {\bibinfo  {journal} {Eur. Phys. J. C}\ }\textbf {\bibinfo {volume} {78}},\ \bibinfo {pages} {65} (\bibinfo {year} {2018})},\ \Eprint {https://arxiv.org/abs/1709.04392} {arXiv:1709.04392 [hep-th]} \BibitemShut {NoStop}%
\bibitem [{\citenamefont {Kastor}\ \emph {et~al.}(2009)\citenamefont {Kastor}, \citenamefont {Ray},\ and\ \citenamefont {Traschen}}]{Kastor:2009wy}%
  \BibitemOpen
  \bibfield  {author} {\bibinfo {author} {\bibfnamefont {D.}~\bibnamefont {Kastor}}, \bibinfo {author} {\bibfnamefont {S.}~\bibnamefont {Ray}},\ and\ \bibinfo {author} {\bibfnamefont {J.}~\bibnamefont {Traschen}},\ }\bibfield  {title} {\bibinfo {title} {{Enthalpy and the Mechanics of AdS Black Holes}},\ }\href {https://doi.org/10.1088/0264-9381/26/19/195011} {\bibfield  {journal} {\bibinfo  {journal} {Class. Quant. Grav.}\ }\textbf {\bibinfo {volume} {26}},\ \bibinfo {pages} {195011} (\bibinfo {year} {2009})},\ \Eprint {https://arxiv.org/abs/0904.2765} {arXiv:0904.2765 [hep-th]} \BibitemShut {NoStop}%
\bibitem [{\citenamefont {Ben~Achour}\ \emph {et~al.}(2026)\citenamefont {Ben~Achour}, \citenamefont {Cisterna},\ and\ \citenamefont {Roussille}}]{BenAchour:2025lkx}%
  \BibitemOpen
  \bibfield  {author} {\bibinfo {author} {\bibfnamefont {J.}~\bibnamefont {Ben~Achour}}, \bibinfo {author} {\bibfnamefont {A.}~\bibnamefont {Cisterna}},\ and\ \bibinfo {author} {\bibfnamefont {H.}~\bibnamefont {Roussille}},\ }\bibfield  {title} {\bibinfo {title} {{A circular Disformal Kerr black hole}},\ }\href {https://doi.org/10.1088/1475-7516/2026/04/041} {\bibfield  {journal} {\bibinfo  {journal} {JCAP}\ }\textbf {\bibinfo {volume} {04}},\ \bibinfo {pages} {041}},\ \Eprint {https://arxiv.org/abs/2512.19549} {arXiv:2512.19549 [gr-qc]} \BibitemShut {NoStop}%
\bibitem [{\citenamefont {Frolov}\ \emph {et~al.}(2017{\natexlab{b}})\citenamefont {Frolov}, \citenamefont {Krtous},\ and\ \citenamefont {Kubiznak}}]{Frolov:2017kze}%
  \BibitemOpen
  \bibfield  {author} {\bibinfo {author} {\bibfnamefont {V.~P.}\ \bibnamefont {Frolov}}, \bibinfo {author} {\bibfnamefont {P.}~\bibnamefont {Krtous}},\ and\ \bibinfo {author} {\bibfnamefont {D.}~\bibnamefont {Kubiznak}},\ }\bibfield  {title} {\bibinfo {title} {{Black holes, hidden symmetries, and complete integrability}},\ }\href {https://doi.org/10.1007/s41114-017-0009-9} {\bibfield  {journal} {\bibinfo  {journal} {Living Rev. Rel.}\ }\textbf {\bibinfo {volume} {20}},\ \bibinfo {pages} {6} (\bibinfo {year} {2017}{\natexlab{b}})},\ \Eprint {https://arxiv.org/abs/1705.05482} {arXiv:1705.05482 [gr-qc]} \BibitemShut {NoStop}%
\end{thebibliography}

%

\end{document}